\documentclass[fleqn,10pt]{SelfArx}
\usepackage{lipsum}

\definecolor{color1}{RGB}{0,0,90} 
\definecolor{color2}{RGB}{0,20,20} 

\usepackage{hyperref} 
\hypersetup{hidelinks,colorlinks,breaklinks=true,urlcolor=color2,citecolor=color1,linkcolor=color1,bookmarksopen=false,pdftitle={Title},pdfauthor={Author}}

 \def\gsim{\lower.4ex\hbox{$\;\buildrel >\over{\scriptstyle\sim}\;$}}

\JournalInfo{Geomagnetism \& Aeronomy, 2022, vol.62, No.7} 
\Archive{}
\PaperTitle{
Inferring Quadrupolar Dynamo Mode from Sunspot Statistics
}
\Authors{L.~L.~Kitchatinov\textsuperscript{1,2}*}
\affiliation{\textsuperscript{1}\textit{Institute of Solar-Terrestrial Physics, Lermontov Str. 126A, Irkutsk, 664033, Russia}}
\affiliation{\textsuperscript{2}\textit{
Pulkovo Astronomical Observatory, Pulkovskoe Sh. 65, St. Petersburg, 196140, Russia
}}
\affiliation{*\textbf{E-mail}: kit@iszf.irk.ru}

\newcommand{\keywordname}{Keywords}

\Abstract{Observations of long-term north-south asymmetry in solar activity demand the equator-symmetric (quadrupolar) mode be present in the solar magnetic field in line with the dominant antisymmetric (dipolar) mode. This paper proposes treating the sunspot area as a proxy for subsurface toroidal magnetic flux to infer the quadrupolar mode of the solar dynamo from sunspot data. Toroidal pseudo-fluxes (PF) in the northern and southern hemispheres are defined as a signed sunspot area with plus or minus sign prescribed to them in accord with the Hale's sunspot polarity rules. Statistical correlation analysis and wavelet analysis of so-defined PFs reveal quadrupolar oscillations with a period of about 16\,yr and amplitude of about 0.17 relative to the amplitude of the dominant 22-yr dipolar mode.}

\begin{document}

\flushbottom 

\maketitle 


\thispagestyle{empty} 

\section{Introduction} 
Solar dynamo models allow two types of magnetic field mo\-des: symmetric (quadrupolar) and antisymmetric (dipolar) about the solar equator (Charbonneau, 2020). Observations point to the dipolar mode as being dominant on the Sun. Hale's polarity laws (Hale et al., 1919) imply opposite signs of the subsurface toroidal field in the northern and southern hemispheres. Polar magnetic fields tend to have opposite polarities as well (see Sect. 3.6 in Hathaway, 2015). Stenflo (1988) analysed 26 years of synoptic observations of the Sun's magnetic field and found that the field of dipolar parity is about ten times stronger compared to the quadrupolar field and the dipolar field only shows a 22-yr magnetic cycle (see also Stenflo and Güdel, 1988).

It is however unlikely that there is no cycling quadrupolar field on the Sun. A purely dipolar field is equatorially symmetric in the sense of magnetic energy. The energy symmetry would mean equal levels of magnetic activity in the northern and southern hemispheres. Observations however show statistically significant long-term north-south (NS) asymmetry varying on a time scale of several solar cycles (Oliver and Ballester, 1994; Zolotova and Ponyavin, 2006; Badalyan and Obridko, 2011; Nagovitsyn and Kuleshova, 2015; Das et al., 2021).

A theory of the long-term hemispheric asymmetry of solar activity has to explain where the large-scale quadrupolar field is coming from. Several theoretical models for NS solar asymmetry have been proposed converging to the idea that the quadrupolar dynamo mode results from equator-asymme\-tric fluctuations in dynamo parameters (Usoskin et al., 2009; Schüssler and Cameron, 2018; Karak et al., 2018; Nepomnyashchikh et al., 2019; Nagy et al., 2019; Kitchatinov and Khlystova, 2021). The models predict quasi-periodic quadru\-polar oscillations with smaller amplitude and somewhat shor\-ter period compared to the basic dipolar mode. Irregular wandering in amplitude and phase is also predicted for the randomly forced quadrupolar mode. The predictions remain to be confronted with observations.

Observational testing of whether quadrupolar magnetic oscillations are pre\-s\-ent in the Sun is however not easy because theory and observations are \lq speaking different langua\-ges'. The dynamo theory operates with vector magnetic fields and observations of solar activity refer to sunspot numbers and areas or other essentially positive scalar quantities (the number of solar cycles covered by magnetographic observations is too small for statistical analysis).

This paper attempts to use the statistics of sunspot areas as a proxy for subsurface magnetic flux to reveal the quadrupolar dynamo mode. Disregarding the small rotationally induced average tilt and large tilt scatter (Jiang et al., 2014), bipolar groups of sunspots tend to be aligned in an east-west direction. The orientation can be understood as the manifestation of a deep subsurface toroidal field (D'Silva and Choudhuri, 1993; Sch\"ussler, 2002). The field strength averaged over the sunspot varies within a factor of about 1.5 only (Solanki, 2003). The spots' magnetic flux is roughly proportional to the area (Nagovitsyn et al., 2016). We define the pseudo-flux (PF) of the toroidal magnetic field separately for the northern and southern hemispheres as a sunspot area with a positive or negative sign prescribed in accord with Hale's polarity rules. Statistical analysis of so-defined PFs shows quasi-periodic quadrupolar oscillations with a period of about 16\,yr (8\,yr for energy oscillations) and about six times smaller mean amplitude compared to the well-known 22-yr dipolar oscillations. This is probably the first observati\-on-based detection of a quadrupolar dynamo mode for the Sun.
\section{Data and method}
We use the joint series of monthly averages of sunspot area data of the Royal Greenwich Observatory and Solar Optical Observing Network of USAF\footnote{\url{https://solarscience.msfc.nasa.gov/greenwch.shtml}}. The series consists of sunspot areas in millionths of a hemisphere (mph) for 1709 months from May 1874 to September 2016 separately for the northern and southern hemispheres. Dating of the series was chan\-ged to measuring time ($t$) in years by the rule $t = Year + (Month\,No - 0.5)/12$, i.e. the difference in month durations was neglected.

\begin{figure}
\centerline{\includegraphics[width=\linewidth]{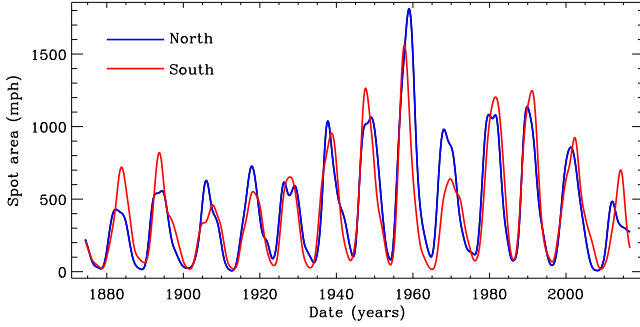}}
\caption{Area of the northern (blue line) and southern (red) spots smoothed with
        the 24-months Gaussian filter of Eq.\,(\ref{1}).}
    \label{f1}
\end{figure}

When it was necessary to smooth noisy monthly data, smoothing with the 24-month Gaussian filter (Hathaway, 2015)
\begin{equation}
    \Phi(t) = \mathrm{e}^{-t^2/2} - \mathrm{e}^{-2} (3 - t^2/2)
    \label{1}
\end{equation}
was applied, where $t$ is time in years. Figure~\ref{f1} shows the sunspot area for the northern and southern hemispheres smo\-othed with this filter. This figure gives another version of Fig.\,35 from (Hathaway, 2015) where the two figures overlap. The new version is used to show that though some cycles have double maxima, all minima fall on a certain date. Dates of the minima in the northern and southern hemispheres  are listed in Tab.\,\ref{t1}.

\begin{table}\centering
\caption{Dates and values of sunspot area minima in the northern and southern hemispheres. The area values are monthly data smoothed with 24-month Gaussian filter of Eq.\,(\ref{1}).}
\label{t1}
\begin{tabular}{ccccc}
  \hline
          & \multicolumn{2}{c}{North} & \multicolumn{2}{c}{South} \\
 Cycle    & Date & Area & Date  & Area  \\
   No      &(Yr/Mon) & (mph) & (Yr/Mon) & (mph) \\ \hline
  12 & 1878/06 & 20.3 & 1878/09 & 22.5 \\
  13 & 1889/02 & 15.6 & 1889/10 & 62.9 \\
  14 & 1901/03 & 23.8 & 1901/09 & 18.4 \\
  15 & 1912/09 & 6.7  & 1913/03 & 19.6 \\
  16 & 1923/05 & 89.9 & 1923/09 & 42.5 \\
  17 & 1934/02 & 62.4 & 1933/02 & 26.6 \\
  18 & 1944/05 & 104.1& 1943/09 & 76.2 \\
  19 & 1954/02 & 81.4 & 1954/02 & 47.2 \\
  20 & 1964/09 & 102.5& 1965/01 & 15.9 \\
  21 & 1976/02 & 117.4& 1976/03 & 74.8 \\
  22 & 1986/02 & 75.4 & 1986/04 & 81.5 \\
  23 & 1996/05 & 43.6 & 1996/05 & 67.3 \\
  24 & 2008/03 & 7.4  & 2009/02 & 18.3 \\
\hline
\end{tabular}
\end{table}

We consider the spot area for a given hemisphere as proxy for the subsurface toroidal magnetic flux in this hemisphere and introduce the toroidal PF by prescribing a plus or minus sign to the monthly spot area in a given hemisphere in accord with the Hale's polarity rule. This rule means a positive (negative) PF in the northern (southern) hemisphere for solar cycle 24 and the sign reversal at each minimum date of  Tab.\,\ref{t1} for each alternate cycle in either hemisphere.
This rule leaves the sign of PFs at the dates of spot area minima uncertain. The signs of PFs for the two neighboring dates are however definite. The same sign was prescribed to the PF of a minimum date as the neighboring PF with larger absolute value has. The sign definition for small PFs of the minima dates is of minor significance. Nullifying PFs for the minima dates leads to essentially the same results.

The dipolar ($FD$) and quadrupolar ($FQ$) parts of the total PF can be compiled from their northern ($FN$) and southern ($FS$) parts as follows
\begin{eqnarray}
    FQ &=& (FN + FS)/2,
    \nonumber \\
    FD &=& (FN - FS)/2.
    \label{2}
\end{eqnarray}

The aim of this paper is to probe sunspot data for the presence of long-term quadrupolar oscillations in the solar magnetic field. The presence of quasi-periodic variations in a time dependent variable $F(t)$ can be probed with the correlation function
\begin{equation}
    C(n\delta t) = \frac{1}{N - n} \sum_{i=1}^{N-n}F(t_i)F(t_{i+n}),
    \label{3}
\end{equation}
where $\delta t$ is the time increment in the data series and $N$ is the total number of the data points; $\delta t$ is one month and $N = 1709$ for the sunspot data we use. Kitchatinov and Khlystova (2021) used the correlation function to reveal quasi-periodic quadrupolar oscillations in magnetic fields of their dynamo model. If the function $F(t)$ is strictly periodic, the correlation function (\ref{3}) is periodic as well. Phase-wandering of quasi-periodic oscillations in $F(t)$ result in decaying oscillations of the correlation function (Kitchatinov and Khlystova, 2021).

Another method in use for revealing quasi-periodic variations is wavelet analysis. Maxima in the power
\begin{equation}
    W_t(p) = \mid \hat{F}_t(p)\mid ^2
    \label{4}
\end{equation}
of the wavelet transform
\begin{equation}
    \hat{F}_t(p) = \sum_{i=1}^{N} \frac{\delta t}{p}
    \Psi^* \left(\frac{t_i - t}{p}\right)F(t_i)
    \label{5}
\end{equation}
show the dates $t$ where oscillations of the period $p$ in the variable $F(t)$ are most pronounced (see, e.g., Torrence and Campo, 1998). We use wavelet Morlet basis function
\begin{equation}
    \Psi(\tau) = \pi^{-1/4}\mathrm{e}^{-\tau^2/2}
    \left( \mathrm{e}^{-\mathrm{i}\omega\tau} - \mathrm{e}^{-\omega^2/2}\right) ,
    \label{6}
\end{equation}
where $\omega = 2\pi - 1/(4\pi)$ is the wavelet normalized angular frequency. The factor $\delta t/p$ is used in Eq.\,(\ref{5}) instead of conventional $\sqrt{\delta t/p}$ to avoid artificial suppression of short periodicity.
\section{Results and discussion}\label{RD}
\begin{figure}
    \centerline{\includegraphics[width = \linewidth]{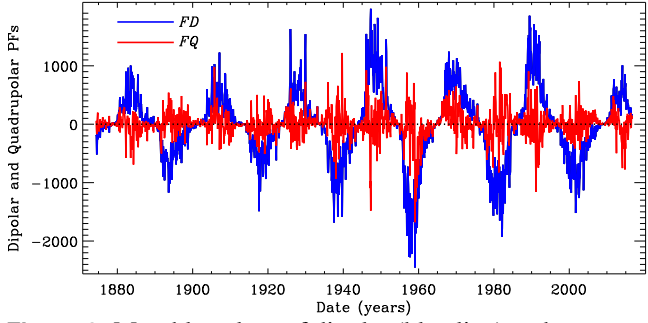}}
    \caption{Monthly values of dipolar (blue line) and quadrupolar (red) parts of PFs of Eq.\,(\ref{2}).}
    \label{f2}
\end{figure}

\begin{figure}
    \centerline{\includegraphics[width = \linewidth]{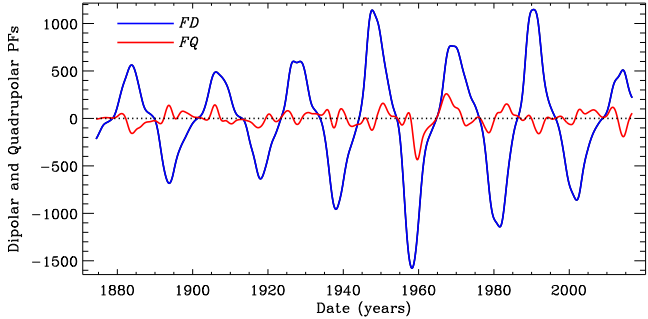}}
    \caption{PFs of dipolar and quadrupolar equatorial parity smoothed with 24-months
             Gaussian filter.}
    \label{f3}
\end{figure}

Dipolar and quadrupolar parts of monthly mean PFs are shown in Fig.\,\ref{f2}. The values are noisy and have to be smoothed. Smoothing with 24-month Gaussian filter of Eq.\,(\ref{1}) reveals cyclic variations in the dominant dipolar PF of Fig.\,\ref{f3}.

Variations of the smoothed quadrupolar PF are less regular as it should be expected for the quadrupolar mode excitation by short-term fluctuations in parameters of the solar dynamo (Sch\"ussler and Cameron, 2018; Nepomnyashchikh et al., 2019). The dynamo model for NS solar asymmetry by Kitchatinov and Khlystova (2021) suggests that the quadrupolar dynamo-mode is excited by the dominant dipolar mode via the equator-symmetric fluctuations in the Babcock-Leigh\-ton mechanism. The model predicts that the quadrupolar magnetic field tends to vary in-phase or in antiphase with the dipolar mode with relatively fast transitions between these two states. The antiphase variation can be seen at the beginning and at the end of the date range of Fig.\,\ref{f3} and in-phase variation is present at the beginning of its second half. Transitions between these two states are not fast however.

\begin{figure}
    \centerline{\includegraphics[width = \linewidth]{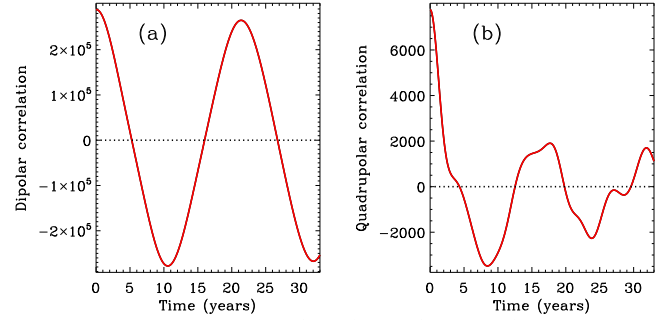}}
    \caption{Autocorrelation functions of Eq.\,(\ref{3}) for the dipolar (a)
             and quadrupolar (b) PFs of Eq.\,(\ref{2}).}
    \label{f4}
\end{figure}

The presence of a periodic part in variations of quadrupolar PF is more evident in its auto-correlation function in Fig.\,\ref{4}. The difference between the instants of successive crossings of zero in the figure estimates half-cycle of the oscillation. Estimation from the right panel of Fig.\,\ref{f4} gives a half-period of about 8 yr for the quadrupolar oscillations. This is distinctly shorter than the 10.7 yrs for the half-period of the dominant dipolar mode estimated from panel (a) in this figure.

\begin{figure}
    \centerline{\includegraphics[width = \linewidth]{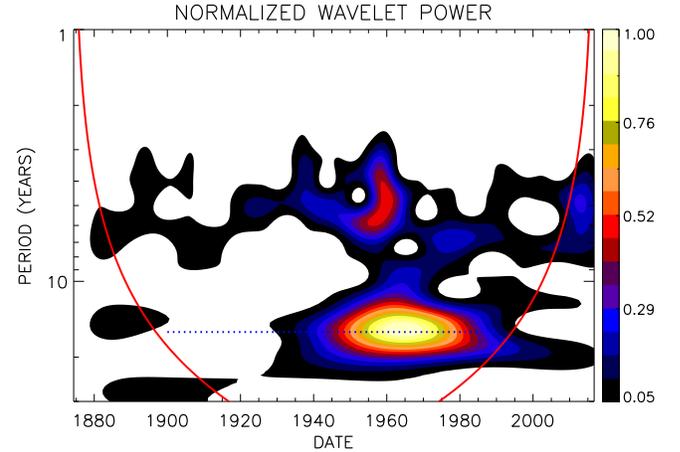}}
    \caption{Morlet wavelet power spectrum for the quadrupolar PF of Fig.\,\ref{f3}
             normalized to the maximum value one. The red line shows the cone of influence (outside of which borders of the data range influence the results). The horizontal dotted line shows the period of 16 yr. Levels below 0.05 are not shown.}
    \label{f5}
\end{figure}

The autocorrelation value at zero argument estimates the mean squared amplitude $A^2 \simeq 2 C(0)$ of a quasi-periodic oscillation. Judging from Fig.\,\ref{f4}, the amplitude of the quadrupolar mode is about 6 times smaller compared to the dipolar mode. This value can be taken as only a rough estimation for the amplitudes of solar dynamo modes because the relation of PFs to the magnetic field strength is not known. The estimated period of the quadrupolar mode is more certain.

\begin{figure}
    \centerline{\includegraphics[width = \linewidth]{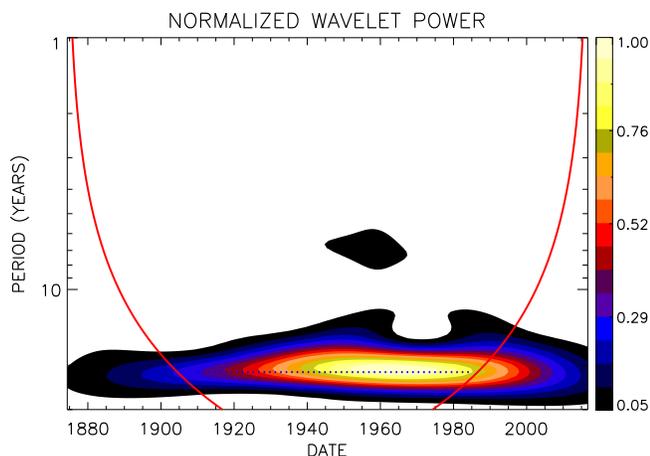}}
    \caption{The same as Fig.\,\ref{f5} but for dipolar PF. The estimated period of 21.4 yr is shown by the dotted line.}
    \label{f6}
\end{figure}

This periodicity is evident in the wavelet power spectrum for the quadrupolar PF of Fig.\,\ref{f5} also. The total period of about 16\,yr is pronounced in the second half of the date range. This is partly because the solar cycles' amplitude is larger for  these dates (Fig.\,\ref{f1}). The known 22\,yr period (21.4\,yr for the present case) of dipolar oscillations is also more pronounced in the second half of the date range of Fig.\,\ref{f6}.

Reproducing the dominance of dipolar magnetic fields was not a trivial task for solar dynamo models (Chatterjee et al., 2004; Hotta and Yokoyama, 2010). Any dynamo model predicts the properties of the subdominant quadrupolar modes also. The 16\,yr period (about 8\,yr for the more commonly used half-period of energy oscillations) of the quadrupolar mode is a further constraint for the solar dynamo models.
\section{Conclusion}
Using the sunspot area as proxy for the magnetic flux of the subsurface toroidal field helps to reveal a subdominant quadrupolar mode in the solar magnetic flux in line with the known dominant dipolar mode. The sign of the so-defined pseudo-flux can be prescribed after the Hale's (Hale et al. (1919) polarity rules. Statistical analysis of the quadrupolar pseudo-flux shows mean period of about 16\,yr in its temporal variations and amplitude of about 0.17 relative to the dipolar mode. The temporal variations are quasi-periodic with irregular variations present in their phase and amplitude, as it should be if the quadrupolar mode is excited by the dominant dipolar mode via random fluctuations in dynamo parameters (Kitchatinov and Khlystova, 2021). Observations of long-term NS asymmetry of solar activity (Oliver and Ballester, 1994; Zolotova and Ponyavin, 2006; Badalyan and Obridko, 2011; Das et al., 2021) imply the presence of a quadrupolar component in the large-scale magnetic field. The presence of a quadrupolar dynamo-mode in the Sun is now confirmed with sunspot data.
\phantomsection
\section*{Acknowledgment}
This work was financially supported by the Ministry of Science and High
Education of the Russian Federation.
\phantomsection
\section*{References}
\begin{description}
\item{} Badalyan, O.\,G. and Obridko, V.\,N., 2011, North-south asymmetry of the sunspot indices and its quasi-biennial oscillations, {\it New Astron.}, vol.\,16, pp.\,357-365.
\item{} Charbonneau, P., 2020, Dynamo models of the solar cycle, {\it Living Reviews in Solar Physics},  vol.\,17, id\,4.
\item{} Chatterjee, P., Nandy, D., and Choudhuri, A.\,R., 2004, Full-sphere simulations of a circulation-dominated solar dynamo: Exploring the parity issue, {\it Astron. Astrophys.}, vol.\,427, pp.\,1019-1030.
\item{} Das, R., Ghosh, A., and Karak, B.\,B., 2022, Is the hemispheric asymmetry of monthly sunspot area an irregular process with long-term memory? {\it Mon. Not. Roy. Astron. Soc.}, vol.\,511, pp.\,472-479.
\item{} D'Silva, S. and Choudhuri, A.\,R., 1993, A theoretical model for tilts of bipolar magnetic regions, {\it Astron. Astrophys.}, vol.\,272, pp.\,621-633.
\item{} Halle, G.\,E., Ellerman, F., Nicholson, S.\,B., and Joy, A.\,H., 1919, The magnetic polarity of sun-spots. {\it Astrophys. J.}, vol.\,49, pp.\,153-178.
\item{} Hathaway, D.\,H., 2015, The solar cycle, {\it Living Reviews in Solar Physics}, vol.\,12, id\,4.
\item{} Hotta, H. and Yokoyama, T., 2010, Solar parity issue with flux-transport dynamo, {\it Astrophys. J. Lett.}, vol.\,714, pp.\,308-312.
\item{} Jiang, J., Cameron, R.\,H., and Sch\"ussler, M., 2014, Effects of the scatter in sunspot group tilt angles on the large-scale magnetic field at the solar surface, {\it Astrophys. J.}, vol.\,791, id\,5.
\item{} Karak, B.\,B., Mandal, S., and Banerjee, D., 2018, Double peaks of the solar cycle: An explanation from a dynamo model, {\it Astrophys. J.}, vol.\,866, id\,17.
\item{} Kitchatinov, L. and Khlystova, A., 2021, Dynamo model for north-south asymmetry of solar activity, {\it Astrophys. J.}, vol.919, id\,36.
\item{} Nagovitsyn, Yu.\,A. and Kuleshova, A.\,I., 2015, North-south asymmetry of solar activity on a long timescale, {\it Geomagnetism and Aeronomy}, vol.55, pp.\,887-891.
\item{} Nagovitsyn, Yu.\,A., Tlatov, A.\,G., and Nagovitsyna, E.\,Yu., 2016, The area and absolute magnetic flux of sunspots over the past 400 years, {\it Astronomy Reports}, vol.60, pp.\,831-838.
\item{} Nagy, M., Lemerle, A., and Charbonneau, P., 2019, Impact of rogue active regions on hemispheric asymmetry, {\it Advances in Space Research}, vol.\,63, pp.\,1425-1433.
\item{} Nepomnyashchikh, A., Mandal, S., Banerjee, D., and Kitchatinov, L., 2019, Can the long-term hemispheric asymmetry of solar activity result from fluctuations in dynamo parameters? {\it Astron. Astrophys.}, vol.\,625, id\,A37.
\item{} Oliver, R. and Ballester, J.\,L., 1994, The north-south asymmetry of sunspot areas during solar cycle 22, {\it Solar Phys.}, vol.\,152, pp.\,481-485.
\item{} Sch\"ussler, M. and Cameron, R.\,H., 2018, Origin of the hemispheric asymmetry of solar activity, {\it Astron. Astrophys.}, vol.\,618, id\,A89.
\item{} Solanki, S.\,K., 2003, Sunspots: An overview, {\it Astron. Astrophys. Rev.}, vol.\,11, pp.\,153-286.
\item{} Stenflo, J.\,O., 1988, Global wave patterns in the Sun's magnetic field, {\it Astrophys. Space Sci.}, vol.\,144, pp.\,321-336.
\newpage
\item{} Stenflo, J.\,O. and G\"udel, M., 1988, Evolution of solar magnetic fields - Modal structure, {\it Astron. Astrophys.}, vol.\,191, pp.\,137-148.
\item{} Torrence, C., Compo, G.\,P., 1998, A practical guide to wavelet analysis, {\it Bulletin of the American Meteorological Society.}, vol.\,79, pp.\,61-78.
\item{} Usoskin, I.\,G., Sokoloff, D., Moss, D., 2009, Grand minima of solar activity and the mean-field dynamo, {\it Solar Phys.}, vol.\,254, pp.\,345-355.
\item{} Zolotova, N.\,V. and Ponyavin, D.\,I., 2006, Phase asynchrony of the north-south sunspot activity, {\it Astron. Astrophys.}, vol.\,449, pp.\,L1-4.
\end{description}
\end{document}